\documentstyle[12pt]{article}

\newcommand{\sect}[1]{\setcounter{equation}{0}\section{#1}}

\newtheorem{proposition}{Proposition}

\def\be{\begin{equation}}
\def\ee{\end{equation}}
\def\bea{\begin{eqnarray}}
\def\eea{\end{eqnarray}}

\def\casi{$(\surd)$}
\def\si{$\surd$} 

\def\la{\langle}
\def\ra{\rangle}

\def\ei{\epsilon_1}
\def\eii{\epsilon_2}
\def\ve{\varepsilon}

\def\k{\kappa} 
\def\m{\mu} 
\def\Ck{{\rm{\ \!C\/}}}       
\def\Sk{{\rm{\ \!S\/}}}

\def\>#1{{\bf #1}}                 
\def\1{\'{\i}}                          
\def\R{\rm I\kern-.2em R}
\def\back{\!\!\!\!\!\!}

\def\JJ{{J}}
\def\NN{{N}}
\def\z{{{z}}}

\def\Jt{\JJ_3}
\def\Nt{\NN_3}
\def\Jpm{\JJ_\pm}
\def\Npm{\NN_\pm}
\def\Nmp{\NN_\mp}
\def\Jp{\JJ_+}
\def\Np{\NN_+}
\def\Jm{\JJ_-}
\def\Nm{\NN_-}
\def\Ja{\JJ_m}
\def\Na{\NN_m}

\def\tfrac#1#2{{\scriptstyle{\frac{#1}{#2}}}}         

\begin{document}

\thispagestyle{empty}

\hfill\today 

\vspace{2cm}

\begin{center}
{\LARGE{\bf{Non-standard quantum   so(2,2) and beyond}}}
\end{center}

\bigskip\bigskip\bigskip

\begin{center}
A. Ballesteros$^{\dagger}$, F.J. Herranz, M.A. del Olmo and M. Santander
\end{center}

\begin{center}
{\it  {Departamento de F\1sica Te\'orica, Universidad de
Valladolid}\\  E-47011, Valladolid, Spain}
\end{center}

\begin{center} {\it  { {}$^{\dagger}$ Departamento de F\1sica, 
Universidad de
Burgos}\\ E.U. Polit\'ecnica,  E-09006, Burgos, Spain} \end{center}

\bigskip\bigskip\bigskip

\begin{abstract} 
A new ``non-standard" quantization of the universal enveloping algebra
of the split (natural)  real form $so(2,2)$ of $D_2$ is presented. Some
(classical) graded contractions of $so(2,2)$ associated to a $\>Z_2
\times  \>Z_2$ grading are studied, and the automorphisms defining this
grading are generalized to the quantum case, thus providing quantum
contractions of this algebra. This produces a new family of
``non-standard" quantum algebras. Some of these algebras can be realized
as (2+1) kinematical algebras; we explicitly introduce a new deformation of
Poincar\'e algebra, which is naturally linked to the null plane basis. 
Another
realization of these quantum algebras as deformations of the conformal
algebras for the two-dimensional Euclidean, Galilei and Minkowski spaces
is given, and its new properties are emphasized. 
\end{abstract}

\newpage 

\sect {Introduction} 

Quantum algebras can be understood as deformations of Lie bialgebras. 
For a semisimple Lie algebra $g$, all its Lie bialgebras  are coboundary 
structures coming from classical $r$--matrices. Thus, the classification
of $r$--matrices  provides a first description of the inequivalent
quantizations of $g$. In general, $r$--matrices can be either {\it
non-degenerate} (i.e., a skew solution of the modified Yang--Baxter
equation (MYBE)) or {\it degenerate} (i.e., a skew solution of the
classical Yang--Baxter equation (CYBE)) \cite{PCh}. All  non-degenerate
$r$--matrices for simple Lie algebras were obtained by Belavin and
Drinfel'd \cite{BDr}.

For $sl(2,\R)$ there exist three classes of Lie bialgebras  \cite{PCh}:
one of them ($r=2 \Jp \wedge \Jm $) is non-degenerate and corresponds to
the well-known Drinfel'd--Jimbo quantum deformation, hereafter called
{\it standard} deformation. The two  remaining solutions are generated by
the trivial one $r=0$ and  by $r= \Jt \wedge \Jp $, respectively. The
quantization of the latter has been recently worked out by Ohn
\cite{Ohn}, and provides a new kind of {\it non-standard} quantum
deformation of $sl(2,\R)$.

Many efforts have been also devoted to the obtention of quantum
non-semisimple algebras from contraction of the standard deformation, but
--to our knowledge-- no contractions of this non-standard quantum
algebra $U_{z}sl(2,\R)$ or related algebras have been explored so far. The
aim of this paper is twofold: firstly, to construct such a non-standard
quantum $so(2,2)$ by using the prescription  
$U_zso(2,2) \simeq U_zso(2,1)\oplus
U_{-z}so(2,1)$ \cite{CGST1,CGST2} as applied to the Ohn non-standard
quantization \cite{Ohn} for the two $so(2,1) \simeq sl(2,\R)$ copies, and to
compare it with the standard quantum $so(2,2)$. Secondly, to study the
quantum contractions of the standard and non-standard quantum $so(2,2)$
algebras. In particular, new ``non-standard" quantum kinematical (de
Sitter, anti-de Sitter and Poincar\'e) algebras in (2+1) dimensions are
obtained, as well as new quantum conformal algebras in 2
dimensions.

The usual way to deal with contractions is the In\"on\"u--Wigner (IW)
\cite{IW} scheme, where a single parameter is made to go towards some
singular limit. Recently, a more comprehensive contraction method based on
the preservation of a given grading of the Lie algebra has been developed
by Moody and Patera \cite{PatMoo}; this method includes both the IW
contractions as well as some kind of general ``Weyl unitary trick", which
relates different real forms of the same complex  algebra. We adopt in
this paper this graded contraction point of view, and we consistently
consider only real forms of Lie algebras. Therefore, we start in
Section 2 with a (real) grading of the natural non-compact real
form $so(2,2)$ determined by a set of commuting involutive automorphisms. A
family of graded contracted algebras depending on three real parameters,
$g_{(\m_1, \m_2, \m_3)}$, is naturally distinguished, and includes the
algebra $so(2,2)$ as $g_{(+1, +1, +1)}$. The information conveyed by these
automorphisms is relevant in order to classify the quantum deformations of
the graded contractions of $so(2,2)$.

In principle, we could try to quantize these graded contractions by using
the two different (\it standard \rm and \it non-standard\rm) $so(2,2)$
deformations which are presented in Section 3. Furthermore, as it is shown
in Section 4, it turns out that in each case not all $U_z so(2,2)$
contractions are allowed, but only some well-defined family.   Some of the
results obtained contracting the standard deformation to the family $g_{(\m_1,
+1, \m_3)}$ can be related to those already known 
(see e.g. \cite{BHOS3}), but
we get a new set of non-standard deformations for the family of algebras
$g_{(\m_1, \m_2, +1)}$ which include new deformations of the most interesting
(2+1) kinematical algebras: De Sitter and Poincar\'e algebras. 

When these algebras are understood as conformal algebras in two
dimensions, we get new quantum non-standard conformal algebras,  which,
like the standard ones, contains Hopf subalgebras generated by
isometries of the space augmented just with dilations. However, these
quantum algebras have completely new features as compared with the known
ones (see e.g. \cite{Jus}). In particular, both translations appear in
a completely symmetrical way, and both of them are \it primitive\rm. All
these results are presented in Section 5.


\section  {Graded contractions of so(2,2)}

Let us recall briefly the theory of graded contractions of Lie algebras
\cite{PatMoo}. Suppose $L$ is a \it real \rm Lie algebra, graded by an
Abelian finite group $\Gamma$ whose product is denoted additively. The
grading is a decomposition of the vector space structure of $L$ as 
\be
L=\bigoplus_{\mu\in \Gamma} L_\mu ,  \label{gradingdef}
\ee
such that for $x \in L_\mu$ and $y \in L_\nu$  then $[x,y]$ belongs to
$L_{\mu+\nu}$. This is written as:
\be
[L_\mu,L_\nu]\subseteq L_{\mu+\nu} , \qquad
\mu, \, \nu, \, \mu+\nu\in \Gamma.  \label{condicgc}
\ee
A \it graded contraction \rm of the Lie algebra $L$ is a Lie algebra
$L^{(\ve)}$  with the same vector space structure as $L$, but Lie brackets
for $x\in L_\mu$, $y\in L_\nu$ modified as:
\be
[x,y]_\ve := \ve_{\mu,\nu} \,[x,y], \label{defgc}
\ee
where the {\it contraction parameters} $\ve_{\mu,\nu}$ are real numbers
such that $L^{(\ve)}$ is indeed a Lie algebra; this implies that they
should satisfy the {\it contraction equations}:
\bea
&&\ve_{\mu,\nu} = \ve_{\nu,\mu} \nonumber\\ 
&&\ve_{\mu,\nu} \, \ve_{\mu+\nu,\sigma}= \ve_{\mu,\nu+\sigma} \,
\ve_{\nu,\sigma}
\label{equationsgc}
\eea
for all \it relevant \rm values of indices. Each set of parameters $\ve$
which is a solution of (\ref{equationsgc}) defines a contraction; two
contractions $\ve^{(1)}$, $\ve^{(2)}$ are equivalent if they are related
by: 
\be
\ve^{(2)}_{\mu,\nu}=\ve^{(1)}_{\mu,\nu} \, \frac{r_\mu r_\nu}{r_{\mu+\nu}}, 
\label{equivgc}
\ee
(without summation over repeated indices) where the $r$'s are {\it
non-zero} real numbers which should be thought of as {\it scaling
factors} of the grading subspaces.

Even if the contraction parameters associated to {\it any} pair of
elements $\mu,\nu$ in $\Gamma$ seem to appear in the equations
(\ref{equationsgc}), many of them will not, because it could happen that
in the non-contracted algebra, all the elements $x \in L_\mu$ commute
with the elements $y \in L_\nu$; the parameters $\ve_{\mu,\nu}$
corresponding to $[L_\mu, L_\nu]=0$ are irrelevant and the equations
(\ref{equationsgc}) which contain such parameters do not appear.

Let us consider two copies of $so(2,1)$, each one with basis
$\{\Jt^l,\Jpm^l\}$  $(l=1,2)$, and commutation relations given by
\be
[\Jt^l,\Jpm^l]=\pm 2 \Jpm^l,\qquad [\Jp^l,\Jm ^l] = \Jt^l. 
\label{su2}
\ee

The set of generators  $\{\Jt,\Jpm,\Nt,\Npm\}$ defined by
\be
\Ja=\Ja^1+\Ja^2,\quad \Na=\Ja^1-\Ja^2,\quad m=+,-,3;\label{generadoresso4}
\ee   
closes a $so(2,2)$ Lie algebra with commutation rules
\bea
&&[\Jt,\Jpm]=[\Nt,\Npm]=\pm 2 \Jpm,\nonumber\\
&&[\Jt,\Npm]=[\Nt,\Jpm]=\pm 2 \Npm,\label{algebraso4}\\
&&[\Jp ,\Jm ]=[\Np ,\Nm ]=\Jt,\nonumber\\
&&[\Jpm,\Nmp]=\pm \Nt,\quad [\Ja,\Na]=0,\ \ m=+,-,3.\nonumber
\eea

The two second order Casimirs for this algebra are:
\bea
&&{\cal C}_1=\frac 12 \Jt^2 +\frac 12 \Nt^2 + \Jp  \Jm  + \Jm  \Jp  + \Np 
\Nm  + \Nm 
\Np ,\label{casimirso4i}\\ 
&&{\cal C}_2=\frac 12 \Jt \Nt + \Jp  \Nm  +\Jm  \Np  .\label{casimirso4ii}
\eea

The Lie algebra mappings defined by:
\be  
S^{(\ei,\eii)} : 
(\Jt,\Jpm,\Npm,\Nt) \to (\Jt, \eii \Jpm, \ei\eii \Npm, \ei \Nt),
\label{involutionsso4}
\ee
where $\ei,\ \eii \in \{ 1, -1\}$, are four commuting involutive
automorphisms of $so(2,2)$ which generate a $\>Z_2 \times \> Z_2$ grading
of the $so(2,2)$ Lie algebra. In particular
$S^{(+,+)}= 1$, 
and sometimes we will denote 
$S^{(-,+)} \equiv S_1$, 
$S^{(+,-)} \equiv S_2$ and
$S^{(-,-)} \equiv S_3$. 
With the usual notation for the grading subspaces
according as their elements are either invariant or antiinvariant under
$S_1$ and $S_2$, we have:
\be 
L_{00}=\la \Jt \ra, \quad 
L_{01}=\la \Jp , \Jm  \ra, \quad 
L_{10}=\la \Nt \ra, \quad 
L_{11}=\la \Np , \Nm  \ra. \quad 
\label{gradingso4}
\ee
The contraction coefficients $\varepsilon_{00,00}$, $\varepsilon_{00,10}
= \varepsilon_{10,00}$, and $\varepsilon_{10,10}$ are irrelevant, as the
associated pairs of grading subspaces already commute. The remaining
coefficients 
\bea
&&  \gamma=\varepsilon_{00,01},  \quad \chi=\varepsilon_{00,11}, 
\nonumber \\ 
&& \alpha=\varepsilon_{01,01},  \quad \beta=\varepsilon_{11,11}, \quad
\delta=\varepsilon_{01,11},  \quad \xi=\varepsilon_{01,10}, \quad
\tau=\varepsilon_{11,10},   \nonumber
\eea
should satisfy the contraction equations (\ref{equationsgc}):
\be
(\gamma-\chi) \xi = (\gamma-\chi) \tau =(\gamma-\chi) \delta=0, \qquad 
\alpha \chi = \xi \delta, \quad \delta \tau = \gamma \beta, \quad 
\xi \beta=\alpha\tau. \label{GradEq}
\nonumber
\ee  
A naturally distinguished set of solutions of these equations is obtained
by requiring $\gamma=\chi\neq 0$; each such a solution is equivalent
to a solution  with $\gamma=\chi=1$; the {\it general} solution of
this special case can be expressed in terms of three real constants,
$\mu_1, \mu_2, \mu_3$, by means of:
\bea
\alpha= \mu_2 \mu_3, \quad 
\beta =  \mu_1\mu_2, \quad 
\tau  =       \mu_1, \quad  \xi=\mu_3, \quad     \delta=\mu_2.
\label{solsGradEq}
\eea
The contracted algebras will be denoted by $g_{(\m_1,\m_2,\m_3)}$. Different
choices of the constants may correspond to equivalent graded contractions
(and therefore to isomorphic algebras). It is a simple exercise to check
that, first, the graded contraction $g_{(\m_1,\m_2,\m_3)}$ is equivalent to
$g_{(-\m_1,-\m_2,-\m_3)}$; the scale factors carrying out this equivalence
correspond to the reversal of $\Nt$, and second, any solution
$g_{(\m_1,\m_2,\m_3)}$ is equivalent to one where each
$\m_1, \m_2, \m_3$ can take on the values $\{+1, 0, -1\}$.

Therefore, the equivalence classes of graded contractions with
$\gamma=\chi\neq 0$ can be represented as the vertices, middle points of
edges, middle points of faces, and centre of a cube with ``antipodal"
identification. This means a total of 14 classes of non-equivalent graded
contractions, which are depicted in Fig. 1, and explicitly listed in Table
I below. 

The most relevant algebras in this list   are $so(2,2)$,
$iso(2,1)$ and $so(3,1)$; each of them appears several times, and can be
either interpreted as the algebras of isometries of the (2+1) anti-de
Sitter, Minkowski and de Sitter spaces, or, alternatively, as
conformal algebras of (1+1) Minkowski and Galilean planes, and of 2d
Euclidean plane.  In order to highlight this last interpretation, and
to distinguish at the same time each of the graded contraction
algebras $g_{(\m_1,\m_2,\m_3)}$ from the $so(2,2)$ we started with,
we choose a new naming $J, P_1, P_2, C_1, C_2, D$ for the generators
of $g_{(\m_1,\m_2,\m_3)}$, in such a way that for
$g_{(1,1,1)}$ they are related to the ones of $so(2,2)$ by:
\be 
   J= \frac{1}{2} N_3,\ P_1,=J_+,\ P_2=N_+,\ 
   C_1=-J_-,\ C_2=N_-,\  D=\frac{1}{2}J_3.\label{basis}
\ee 
The commutation relations and Casimirs of $g_{(\m_1,\m_2,\m_3)}$ 
in this basis are: 
\bea   
&& [J ,P_1]=\m_3 P_2, \quad   [J,P_2]=  \m_1 P_1, \quad 
   [P_1,P_2]=0,       \quad   [D,P_i]=P_i,        \nonumber\\  
&& [J ,C_1]=\m_3 C_2, \quad   [J,C_2]=  \m_1 C_1, \quad
   [C_1,C_2]=0,       \quad   [D,C_i]=-C_i,        \nonumber\\   
&& [P_1 ,C_1]=-2\m_2\m_3 D, \qquad  [P_1 ,C_2]=2\m_2J,  \qquad [D,J]=0,
\label{algebragc}\\  
&& [P_2,C_1]= - 2\m_2 J,    \qquad \quad  [P_2,C_2]= 2\m_1 \m_2 D;
\nonumber
\eea
\bea  
&& \back \back 
{\cal C}_1=  \m_2  J^2 +\m_1\m_2\m_3 D^2 -
     \frac 12 \m_1 (P_1C_1+C_1P_1) + \frac 12 \m_3 (P_2C_2 +C_2P_2),  
\label{casimirgci} \\  
&& \back \back 
{\cal C}_2= \m_2 J D +\frac 12 ( P_1C_2-C_1P_2).  \label{casimirgcii}
\eea

It is therefore clear that $g_{(1,1,1)}$ and $g_{(-1,1,1)}$ are
the conformal algebras of (1+1) Minkowski space and the algebra of the
group of M\"obius transformations in the Euclidean plane; the names of
the generators have been chosen to underline this fact. 
      
The involutive automorphisms $S_i$  act  on the generators of 
$g_{(\m_1,\m_2,\m_3)}$ as:  
\bea  
&& S_1 (D, P_1, C_1, P_2, C_2, J)= (D, P_1, C_1,-P_2,-C_2,-J), 
\nonumber\\  
&& S_2 (D, P_1, C_1, P_2, C_2, J)= (D,-P_1,-C_1,-P_2,-C_2, J),
\label{commautso4}\\  
&& S_3 (D, P_1, C_1, P_2, C_2, J)= (D,-P_1,-C_1, P_2, C_2,-J), 
\nonumber
\eea
and have associated IW contractions, $\iota_i$,
as the limit $\lambda_{i} \to 0$ of the Lie algebra automorphisms: 
\bea   
&&\iota_1 (D, P_1, C_1, P_2, C_2, J) : = 
      (D, P_1, C_1, \lambda_{1} P_2, \lambda_{1}C_2, \lambda _{1}J),  
\nonumber\\    
&&\iota_2 (D, P_1, C_1, P_2, C_2, J) : = 
 (D, \lambda_{2}P_1, \lambda_{2}C_1, \lambda_{2}P_2, \lambda_{2}C_2, J), 
\label{iwcontrac}\\  &&\iota_3 (D, P_1, C_1, P_2, C_2, J) : = 
      (D, \lambda_{3} P_1, \lambda_{3} C_1, P_2, C_2, \lambda_{3} J),
\nonumber
\eea
which appear as the graded contractions $(\mu_1, \mu_2, \mu_3) = (0,1,1),
(1,0,1)$  and $(1,1,0)$. 

The interchange $(P_1, C_1) \leftrightarrow (P_2, C_2)$ is a 
 Lie algebra automorphism
$g_{(\m_1,\m_2,\m_3)} \to g_{(\m_3,\m_2,\m_1)}$, so the list 
of 14 non-equivalent
graded contractions reduces, up to isomorphisms, to 8 Lie algebras. 
Instead
of working with this list, two different choices of representatives
 of the
equivalence classes of graded contractions (named (a) and (b)) will be
useful in this paper: 
\bea
&& \back\back \back \mbox{Type (a):} \ \ g_{(\m_1,+1,\m_3)},
    \ \  \mbox{where } \m_1, \m_3 \in \{+1, 0, -1\}, \\
&& \back\back \back \mbox{Type (a0):} \ \  g_{(\m_1,0,\m_3)},
    \ \  \mbox{where } (\m_1, \m_3) \in 
\{(1,1), (0,1), (-1,1), (1,0), (0,0)\}, \nonumber 
\eea
\bea
&& \back \back \mbox{Type (b):} \quad g_{(\m_1,\m_2,+1)},
    \quad \mbox{where } \m_1,\m_2 \in \{+1, 0, -1\} ,\\
&& \back \back \mbox{Type (b0):} \quad g_{(\m_1,\m_2,0)},
    \quad \mbox{where } (\m_1, \m_2) \in 
    \{(1,1), (0,1), (-1,1), (1,0), (0,0)\}. \nonumber 
\eea 
We shall keep in mind that each of
the families a/a0 or b/b0 contain {\it all} graded contractions up to
equivalence. In Fig. 1, type (a) and type (b) algebras appear
respectively on the top and front faces of the cube. Note that the three
algebras $g_{(\m_1,+1, +1)} (\simeq so(2,2), iso(2,1), so(3,1)$) on the
upper front edge are common to both sets of representatives.

A very concise and practical way to describe this family of graded
contractions of $so(2,2)$ is to get the algebra $g_{(\mu_1,\mu_2,\mu_3)}$
by means of the formal transformation (compare with Man'ko and Gromov,
\cite{Gro}) applied to the $so(2,2) \simeq g_{(+1, +1, +1)}$ generators:
\bea   
&& \back \back\back (J,P_1,P_2,C_1,C_2,D)  =  
\Gamma_{\ }^{(\mu_1,\mu_2,\mu_3)} 
  (N_3/2,J_+,N_+,-J_-,N_-,J_3/2) \cr
&& \back\back :=  
(\sqrt{\mu_1\mu_3}N_3/2,\sqrt{\mu_2\mu_3}J_+,
\sqrt{\mu_1\mu_2}N_+,-\sqrt{\mu_2\mu_3}J_-,\sqrt{\mu_1\mu_2}N_-,J_3/2),
\label{gctrick}
\eea
which is well defined as long as all $\mu_i$ are different from zero, and
where the new generators close the algebra (\ref{algebragc}). This device
has been extensively used by Gromov in a slightly different form which
also uses dual numbers \cite{Gro2}. For our purposes it will suffice to use
this formal replacement when all $\mu_i$ are different from zero, and to
understand (\ref{gctrick}) when some $\mu_i$ goes to zero as the 
corresponding limit
(as in (\ref{iwcontrac})). It should be noted that each IW  contraction
parameter $\lambda_i$ in (\ref{iwcontrac}) corresponds to a factor
$\sqrt{\mu_i}$ in (\ref{gctrick}).

\sect   {Quantum deformations of Uso(2,2)}

\subsection  {Two deformations of Uso(2,1)}

The two quantizations of the non-trivial Lie bialgebras of 
$so(2,1)$ are given by the following statements: 

\begin{proposition}
{\bf (The  standard  quantization of so(2,1))} \cite{Dr} The coproduct
$(\Delta)$, counit $(\epsilon)$, antipode $(\gamma)$ defined by  
\bea
&& \Delta \Jt =1 \otimes \Jt + \Jt\otimes 1,\quad\  
   \Delta \Jpm =e^{-\z\Jt} \otimes \Jpm + \Jpm\otimes e^{\z\Jt};
\label{cosu2s} \\
&& \epsilon(X) =0;\quad \gamma(X)=-e^{{\z}  \Jt}\ X\ e^{-{\z}  
   \Jt},\quad  \mbox{for $X\in \{\Jt,\Jpm\}$},
\label{unsu2s}
\eea
together with the commutation rules 
\be
[\Jt,\Jpm]=\pm 2 \Jpm,\quad  [\Jp ,\Jm]=\frac{\sinh (2\z\Jt)}{2\z},
\label{commsu2s} 
\ee
quantize the $so(2,1)$ Lie bialgebra generated by the classical 
$r$--matrix $r=2 \Jp
\wedge \Jm$ and define the (standard) Hopf algebra $U^{(s)}_\z so(2,1)$.
\end{proposition}

Note   $r$  verifies the MYBE. The center
of $U^{(s)}_\z so(2,1)$ is generated by 
\be
{\cal C}_\z = \frac 12 \cosh 2\z\left(\frac{\sinh (\z \Jt)}{\z}\right)^2+
   \frac{\sinh 2\z}{2\z}(\Jp  \Jm + \Jm \Jp ). \label{casimiru2s}
\ee

\begin{proposition}
{\bf (The  non-standard  quantization of so(2,1))} \cite{Ohn}  
The coproduct, counit, antipode   
\bea
&& \back\!\! \Delta \Jp  =1 \otimes \Jp  + \Jp \otimes 1,\quad 
\Delta \Ja =e^{-\z\Jp } \otimes \Ja + \Ja\otimes
e^{\z\Jp },\quad m=-,3;\label{cosu2n} \\
&&  \back\!\! \epsilon(X) =0;\qquad 
\gamma(X)=-e^{{\z}  \Jp }\ X\ e^{-{\z}  \Jp },\ \ \mbox
{for $X\in \{\Jt,\Jpm\}$},
\label{unsu2n}
\eea
and the commutation relations 
\bea
&& [\Jt,\Jp ]=2\frac{\sinh (\z\Jp )}\z,\qquad  [\Jp ,\Jm]= \Jt,\\
&& [\Jt,J_-]=-J_-\cosh (\z\Jp ) -\cosh (\z\Jp ) \Jm,\quad
\label{commsu2n} 
\eea
define the Hopf algebra $U^{(n)}_\z so(2,1)$ that quantizes the
non-standard  Lie bialgebra structure generated by $r= \Jt \wedge \Jp $.
\end{proposition}

In this case, the $r$--matrix satisfies the CYBE. Now, the center of 
 $U^{(n)}_\z
so(2,1)$ is generated by 
\be
{\cal C}_\z = \frac 1 2 \Jt^2+\frac{\sinh (\z\Jp )}\z \Jm +\Jm \frac{\sinh
    (\z\Jp )}\z+\frac 1 2\cosh^2(\z\Jp ).\label{casimirsu2n}
\ee
 
It is easy to check that the $r$--matrix gives in both cases the first
order terms of the deformation: the antisymmetric part of the first order
of the coproduct of a given generator is just the cocommutator defined by
the corresponding $r$--matrix:
\be
(\Delta - \sigma\circ\Delta)(X)=\delta(X)=
[1\otimes X + X \otimes 1,r].
\label{cocomm}
\ee

\subsection  {Two deformations of Uso(2,2)}

By using the  invariance of $U_zso(2,1)$ under the transformation $\z\to
-\z$, we can write $U^{(m)}_\z so(2,2)=U^{(m)}_\z so(2,1)\oplus
U^{(m)}_{-\z}so(2,1)$ where $m=n$ or $m=s$ according either to the
non-standard or standard $so(2,1)$ deformations. Therefore two
different $q$--deformations of $so(2,2)$ can be obtained in this way. The
proofs  of the propositions 3--8 boils  
down to straightforward checking and will not be given. 

\begin{proposition}{\bf (The  standard  quantization of so(2,2))}. 
The coproduct,
counit, antipode:  
\bea
&& \Delta \Jt=1 \otimes \Jt + \Jt\otimes 1,
   \qquad \Delta \Nt =1 \otimes \Nt + \Nt\otimes 1,\nonumber\\
&& \Delta \Jpm=e^{-\tfrac \z 2 \Nt}\cosh(z \Jt/2) \otimes \Jpm 
   + \Jpm\otimes \cosh(z \Jt/2)e^{\tfrac \z 2 \Nt}\nonumber\\
&& \qquad\quad -e^{- \tfrac \z 2 \Nt}\sinh(z \Jt/2)\otimes
   \Npm  + \Npm\otimes \sinh(z \Jt/2) e^{\tfrac \z
    2 \Nt}, \label{coprodso4qs}\\  
&& \Delta \Npm =e^{-\tfrac \z 2 \Nt}\cosh(z \Jt/2)\otimes 
   \Npm  + \Npm\otimes \cosh(z \Jt/2)e^{\tfrac \z
   2 \Nt}\nonumber\\  
&& \qquad\quad -e^{-\tfrac \z 2 \Nt}\sinh(z \Jt/2)
   \otimes \Jpm  + \Jpm\otimes \sinh(z \Jt/2)e^{\tfrac \z 2
   \Nt};\nonumber
\eea
\be
\epsilon(X) =0; \quad
\gamma(X)=-e^{{\z }  \Nt}\ X\ e^{-{\z}  \Nt},\quad 
\mbox{for  $X\in \{\Jt,\Jpm,\\ \Nt,\Npm\}$};
\label{counAntipodso4qs}
\ee
and the commutation relations 
\bea
&& [\Jt,\Jpm]=[\Nt,\Npm]=\pm 2 \Jpm,\nonumber\\
&& [\Jt,\Npm]=[\Nt,\Jpm]=\pm 2 \Npm,\label{commso4qs}\\
&& [\Jp ,\Jm ]=[\Np,\Nm ]=\frac 1\z \sinh (\z\Jt)\cosh
   (\z\Nt),\nonumber\\ 
&& [\Jpm,\Nmp]=\pm \frac 1\z \sinh (\z\Nt)\cosh
   (\z\Jt) ,\quad [\Ja,\Na]=0,\ \ m=+,-,3, \nonumber
\eea
define the Hopf algebra $U^{(s)}_\z so(2,2)=U^{(s)}_\z so(2,1)\oplus
U^{(s)}_{-\z}so(2,1)$. 
\end{proposition}

The classical $r$--matrix corresponding to this $q$--deformation is 
obtained as the difference of the $r$--matrices generating the two 
$U^{(s)}_\z so(2,1)$ components:
\be
r=r_{1}^{(s)}- r_{2}^{(s)}= 2\Jp ^{1}\wedge \Jm ^{1} -2 \Jp ^{2}\wedge \Jm ^{2}
=\Jp \wedge \Nm  + \Np \wedge \Jm  . \label{ra}
\ee 
This $r$--matrix verifies the MYBE and generates the first order term in $\z $
of the coproduct (\ref{coprodso4qs}).

\begin{proposition}{\bf (The  non-standard  quantization of so(2,2))}. The
coproduct, counit, antipode
\bea
&& \Delta \Jp  =1 \otimes \Jp  + \Jp \otimes 1,
   \qquad \Delta \Np  =1 \otimes \Np  + \Np \otimes 1, \nonumber\\
&&  \Delta \Ja =e^{-\tfrac \z 2 \Np  }\cosh(z \Jp /2 ) \otimes
   \Ja  + \Ja\otimes \cosh(z \Jp /2 ) e^{\tfrac \z 2 \Np }
   \nonumber\\
&& \qquad\quad 
   -e^{-\tfrac \z 2 \Np  }\sinh(z \Jp /2 ) \otimes \Na  + 
    \Na\otimes \sinh(z \Jp /2 ) e^{\tfrac \z 2 \Np},
    \label{coprodso4qn}\\ 
&& \Delta \Na =e^{-\tfrac \z 2 \Np } \cosh(z \Jp /2 )\otimes \Na 
  + \Na\otimes \cosh(z \Jp /2 ) e^{\tfrac \z 2 \Np  } \nonumber\\ 
&& \qquad\quad -e^{-\tfrac \z 2 \Np  }\sinh(z \Jp /2 )\otimes
   \Ja  + \Ja\otimes \sinh(z \Jp /2 )e^{\tfrac \z 2 \Np  },
   \quad m=3,-;\nonumber
\eea
\be
\epsilon(X) =0;\quad
   \gamma(X)=-e^{{\z }  \Np }\ X\ e^{-{\z}  \Np },\quad
   \mbox{for $X\in \{\Jt,\Jpm,\\\Nt,\Npm\}$;} \label{counAntipodso4qn}
\ee
and the commutation relations
\bea
&&\back\back [\Jt,\Jp ] =  \frac 4\z {\sinh(z \Jp /2 )}
  \cosh(z \Np /2 ),\cr 
&&\back\back [\Jt,\Jm ] = -\{\Jm , \cosh(z \Jp /2 ) \cosh(z \Np /2 )\}  
  -\{\Nm ,\sinh(z \Jp /2 )\sinh(z \Np /2 )\} ,\nonumber\\
&&\back\back [\Jt,\Np ] =\frac {4} \z \sinh(z \Np /2 )
   \cosh(z \Jp /2 ), \label{commso4qn}\\
&&\back\back [\Jt,\Nm ] = -\{\Nm, \cosh(z \Jp /2 )\cosh(z \Np /2 )\} 
  -\{\Jm ,\sinh(z \Jp /2 )\sinh(z \Np /2 )\} ,\nonumber\\
&& \back\back[\Nt,\Npm] = [\Jt,\Jpm], \quad 
[\Nt,\Jpm] = [\Jt,\Npm], \nonumber\\
&&\back\back [\Jp ,\Jm ] = [\Np ,\Nm ] = \Jt,\quad  
[\Jpm,\Nmp]=\pm \Nt, \quad
   [\Ja,\Na] = 0, \ \ m=\pm,3, \nonumber 
\eea
(where $\{X,Y\}=XY+YX$ denotes the anticommutator of $X$ and $Y$) define
the Hopf algebra
$U^{(n)}_\z so(2,2)=U^{(n)}_\z so(2,1) \oplus U^{(n)}_{-\z} so(2,1)$.
\end{proposition}

The classical $r$--matrix associated to $U^{(n)}_\z so(2,2)$ is:
\be 
r = \Jt^{1}\wedge \Jp ^{1}  - \Jt^{2}\wedge \Jp ^{2} = 
    \frac 12(\Jt\wedge \Np   + \Nt\wedge \Jp ). \label{rb}
\ee
This is a (skew) solution of the CYBE. The cocommutator (\ref{cocomm})
defined here by (\ref{rb}) is consistent with the coproduct
(\ref{coprodso4qn}) . Note that the $r$--matrices (\ref{ra})  and
(\ref{rb})  are non-degenerate and degenerate, respectively, and preserve
the original character of their components as far as the Yang--Baxter
equation is concerned.

\sect  {Quantum contractions}

The aim of this section is to use both the standard and the non-standard
quantum deformations $U^{(m)}_\z so(2,2)$ $(m=n, s)$ we have just described
in order to obtain quantum deformations  of the graded contractions of
$so(2,2)$ studied in section 2. We first extend the definitions of
classical involutions (\ref{involutionsso4}) and contractions
(\ref{gctrick}) to the quantum case by assuming that they act on the
algebra generators as in the classical case, and that their behaviour on
$\z$ is determined in each case in such a way that the exponents in 
$e^{-\tfrac \z 2 \Nt }$ (see (\ref{coprodso4qs})) or $e^{-\tfrac \z 2
\Np  }$ (see (\ref{coprodso4qn})) are invariant under these 
$q$--involutions and contractions. 

Explicitly, this means that the classical expressions
(\ref{involutionsso4}) and (\ref{gctrick}) should be augmented 
to (with $w$ as 
the contracted deformation parameter):
\bea   
&& \back\back S^{(\ei,\eii)}_{q(s)} :  (D, P_1, C_1, P_2, C_2, J ; \z) \to  
 (D, \eii P_1. \eii C_1, \ei P_2, \ei C_2, \ei\eii J; \ei \z),
\label{qinvolutionss}\\ 
&&\back\back\back   
  (J,P_1,P_2,C_1,C_2,D;w)  = 
 \Gamma^{(\mu_1,\mu_2,\mu_3)}_{q(s)}  (N_3/2,J_+,N_+,-J_-,N_-,J_3/2;z) \cr
&&\back \back\back\!\!  :=  
(\sqrt{\mu_1\mu_3}N_3/2,\sqrt{\mu_2\mu_3}J_+,
\sqrt{\mu_1\mu_2}N_+,-\sqrt{\mu_2\mu_3}J_-,\sqrt{\mu_1\mu_2}N_-,J_3/2;z/
\sqrt{\mu_1\mu_3}), \label{qgctricks} 
\eea
for the standard deformation, and to: 
\bea    
&&\back\back  S^{(\ei,\eii)}_{q(n)} :  (D, P_1, C_1, P_2, C_2, J ; \z) \to  
 (D, \eii P_1, \eii C_1, \ei P_2, \ei C_2, \ei\eii J; \eii \z), 
\label{qinvolutionsn}\\
&&  \back \back\back (J,P_1,P_2,C_1,C_2,D;w)  = 
 \Gamma^{(\mu_1,\mu_2,\mu_3)}_{q(n)}  (N_3/2,J_+,N_+,-J_-,N_-,J_3/2;z) \cr
&&\back \back\back\!\! :=  
(\sqrt{\mu_1\mu_3}N_3/2,\sqrt{\mu_2\mu_3}J_+,
\sqrt{\mu_1\mu_2}N_+,-\sqrt{\mu_2\mu_3}J_-,\sqrt{\mu_1\mu_2}N_-,J_3/2;z/
\sqrt{\mu_1\mu_2}), \label{qgctrickn}
\eea
for the non-standard one. We have in both cases a $\>Z_2 \times \>Z_2$
Abelian group of $q$--involutions. The $q$--deformed Hopf algebras
corresponding to the classical contractions (\ref{algebragc}) of $so(2,2)$
can be directly obtained by applying the transformations
(\ref{qgctricks}) (resp. (\ref{qgctrickn})) to
(\ref{coprodso4qs}--\ref{commso4qs}) (resp. to
(\ref{coprodso4qn}--\ref{commso4qn})). Due to the classical origin
of these transformations, the $q$--deformed commutation relations are
always well defined after (\ref{qgctricks}) (resp. (\ref{qgctrickn})) has
been applied. This does not happen neither for the coproduct nor for the
$r$--matrix, in a different way for each case. Table I  sum up the
results  about the existence and properties of these standard and
non-standard contracted quantum algebras.

\medskip

\noindent
{{\bf Table I.} Existence of standard and non-standard deformations of
the $Z_2 \times Z_2$ graded contracted algebras of $so(2,2)$. (A \si\    
marks the existence of $U_w g$).}
\bigskip

\begin{tabular}{|c|c|c|c|}       
\hline 
Lie Algebra $g_{(\m_1,\m_2,\m_3)}$ & $(\m_1,\m_2,\m_3)$ 
& $U_w^{(s)} g_{(\m_1,\m_2,\m_3)}$ & $U_w^{(n)} g_{(\m_1,\m_2,\m_3)}$\\
\hline
$so(2,2)$   & $(+,+,+)$\quad $(-,-,-)$ & \si & \si \\
$so(2,2)$ & $(+,-,+)$\quad $(-,+,-)$ & \si & \si \\
\hline
$so(3,1)$ & $(+,+,-)$\quad $(-,-,+)$ & \si & \si \\
$so(3,1)$ & $(-,+,+)$\quad $(+,-,-)$ & \si & \si \\
\hline
$iso(2,1)$  & $(0,+,+)$\quad $(0,-,-)$ & \si & \si \\
$iso(2,1)$  & $(+,+,0)$\quad $(-,-,0)$ & \si &     \\
$iso(2,1)$& $(0,-,+)$\quad $(0,+,-)$ & \si & \si \\
$iso(2,1)$& $(+,-,0)$\quad $(-,+,0)$ & \si &     \\
\hline
$t_4(so(1,1)\oplus so(1,1))$  & $(+,0,+)$\quad $(-,0,-)$  
& \casi & \si \\
\hline
$t_4(so(2)\oplus so(1,1))$ & $(-,0,+)$\quad $(+,0,-)$ & \casi &\si \\
\hline
$iiso(1,1)$ & $(0,0,+)$\quad $(0,0,-)$ & \casi & \si \\
$iiso(1,1)$ & $(+,0,0)$\quad $(-,0,0)$ & \casi &     \\
\hline
$i'iso(1,1)$ & $(0,+,0)$\quad $(0,-,0)$ & \si &    \\
\hline
$(\R^4 + \R ) \oplus \R $  & $(0,0,0)$ & \casi &    \\
\hline
\end{tabular}

\newpage

Some remarks are in order:
\begin{itemize}

\item{$U^{(s)}_w g_{(\m_1,+1,\m_3)}$}. We get a well defined Hopf
structure, which comes from an $r$--matrix. This is displayed in Table I 
by a \si. This family is studied in Section 4.1.

\item{$U^{(s)}_w g_{(\m_1,0,\m_3)}$}. In this case the limit
$\m_2 \to 0$ of the transformation (\ref{qgctricks}) applied to the
standard deformation of $so(2,2)$ originates a Hopf algebra which has a
deformed  coproduct and classical (i.e. nondeformed) commutation rules.
This algebra is a deformation of a bialgebra which is not a coboundary:
there is no $r$--matrix for it, as one could expect from the fact that
$wr$ (with $r$ given by (\ref{ra})) diverges when $\m_2$ goes to zero. In
Table I  this is shown by a \casi.  

\item{$U^{(n)}_w g_{(\m_1,\m_2,+1)}$}. We also get a well defined Hopf
structure, which comes from an $r$--matrix. This family is studied in
Section 4.2.

\item{$U^{(n)}_w g_{(\m_1,\m_2,0)}$} In contradistinction with the
standard case, the limit $\m_3 \to 0$ of the transformation
(\ref{qgctrickn}) applied to the non-standard deformation of
$so(2,2)$ {\it does not produce a Hopf algebra}, because the coproduct is
not well-defined in the limit $\m_3 \to 0$.  

\end{itemize}

As all graded contractions of $so(2,2)$ are equivalent to one of type
either (a/a0), or (b/b0), and on the other hand types (a0), (b0) are those
with undefined coproduct or $r$--matrix, we shall only deal with the
quantum deformations arising from the standard family (a):
$g_{(\m_1,+1,\m_3)}$, and from the non-standard family (b):
$g_{(\m_1,\m_2,+1)}$.

\subsection {The standard algebras $U^{(s)}_w g_{(\m_1,+1,\m_3)}$}

\begin{proposition}
{\bf (The standard quantization of $g_{(\m_1,+1,\m_3)}$)} 
When $\m_2=+1$ the transformation (\ref{qgctricks}) gives rise to the
quantum  algebra $U^{(s)}_w g_{(\m_1,+1,\m_3)}$ with $r$--matrix
$r=(P_1\wedge C_2 -P_2\wedge C_1)$ and defined by: 
\bea
&& \Delta J  = 1 \otimes J  + J \otimes 1,
   \qquad \Delta D =1 \otimes D + D\otimes 1,\cr
&& \Delta P_1 =
   e^{-wJ }\Ck_{-\m_1\m_3}(wD ) \otimes P_1  
   + P_1\otimes \Ck_{-\m_1\m_3}(wD) e^{ wJ }\cr 
&& \qquad\quad 
  -e^{-wJ }\Sk_{-\m_1\m_3}(wD) \otimes \m_3 P_2 
  +\m_3 P_2\otimes\Sk_{-\m_1\m_3}(wD) e^{ wJ},\cr 
&& \Delta P_2 =
   e^{-wJ }\Ck_{-\m_1\m_3}(wD ) \otimes P_2  
   + P_2\otimes \Ck_{-\m_1\m_3}(wD) e^{ wJ }\cr 
&& \qquad\quad 
  -e^{-wJ }\Sk_{-\m_1\m_3}(wD) \otimes \m_1 P_1 
  +\m_1 P_1\otimes\Sk_{-\m_1\m_3}(wD) e^{wJ},  \label{coprodgcs}\\ 
&& \Delta C_1 =
   e^{-wJ }\Ck_{-\m_1\m_3}(wD ) \otimes C_1  
   + C_1\otimes \Ck_{-\m_1\m_3}(wD) e^{ wJ }\cr 
&& \qquad\quad 
  +e^{-wJ }\Sk_{-\m_1\m_3}(wD) \otimes \m_3 C_2 
  -\m_3 C_2\otimes\Sk_{-\m_1\m_3}(wD) e^{wJ},\cr 
&& \Delta C_2 =
   e^{-wJ }\Ck_{-\m_1\m_3}(wD ) \otimes C_2  
   + C_2\otimes \Ck_{-\m_1\m_3}(wD) e^{ wJ }\cr 
&& \qquad\quad 
  +e^{-wJ }\Sk_{-\m_1\m_3}(wD) \otimes \m_1 C_1 
  -\m_1 C_1\otimes\Sk_{-\m_1\m_3}(wD) e^{wJ};
  \nonumber 
\eea
\be
\epsilon(X) =0; \qquad
  \gamma(X) = -e^{2{w}  J}\ X\ e^{-2{w} J}, 
  \qquad X\in\{J,P_i,C_i,D\} ; \label{coungcs}
\ee
\bea  
&& [J ,P_1]=\m_3 P_2, \quad [J,P_2]=  \m_1 P_1,\quad [P_1,P_2]=0,
\nonumber\\ 
&& [J ,C_1]=\m_3 C_2, \quad [J,C_2]=  \m_1 C_1,\quad [C_1,C_2]=0,
\nonumber\\  
&& [P_1 ,C_1]=-\frac 1w \m_3 \Sk_{-\m_1\m_3}(2wD)\cosh(2wJ),
\label{commgcs}\\ 
&& [P_2 ,C_2]=\frac 1w \m_1 \Sk_{-\m_1\m_3}(2wD)\cosh(2wJ),\cr
&& [P_1 ,C_2]=[C_1,P_2]=\frac 1w \sinh(2wJ)\Ck_{-\m_1\m_3}(2wD),\cr
&&[D,P_i]=P_i,\quad [D,C_i]=-C_i,\quad [D,J]=0,\quad i=1,2.\nonumber
\eea
\end{proposition} 

We recall that the generalized sine and cosine functions are given by
\cite{BHOS3}
\be
\Ck_{{-\m}}(x)=\frac{e^{\sqrt{{\m}} x}+ e^{-\sqrt{{\m}}
x}}{2}, \qquad  \Sk_{{-\m}}(x)=\frac{e^{\sqrt{{\m}}
x}-e^{-\sqrt{{\m}} x}} {2\sqrt{{\m}}}.
\label{trig3}
\ee

Note that two non-trivial Hopf subalgebras with \it undeformed
\rm commutation brackets are contained in $U^{(s)}_w g_{(\m_1,+1,\m_3)}$:
$\langle J,P_1,P_2,D \rangle$ and $\langle J,C_1,C_2,D\rangle$. 
However, neither $\langle J,P_1,P_2
\rangle$ nor $\langle J,C_1,C_2\rangle$ are Hopf subalgebras;
the generator $D$ is required to close their coproducts.  
This fact  has been noted   in the literature \cite{LN} for higher
dimensional cases   of the algebras $U^{(s)}_w
g_{(\m_1,+1,+1)}$, which can be interpreted as conformal algebras of flat
two-dimensional spaces.

A long but
straightforward computation leads to the next:

\begin{proposition} 
The center of  $U^{(s)}_w g_{(\m_1,+1,\m_3)}$ is generated by
\bea
&& \back\back 
   {\cal{C}}_{1}^q = \frac{\Ck_{-\m_1\m_3}(2w)}{w^2}
   \left[\sinh^2(wJ)\Ck_{-\m_1\m_3}^2(wD)  
   + \m_1\m_3\Sk_{-\m_1\m_3}^2(wD)\cosh^2(wJ)\right] \cr
&& \qquad  
   +\frac{\Sk_{-\m_1\m_3}(2w)}{4w}[ -\m_1 (P_1C_1+C_1P_1)  + 
   \m_3  (P_2C_2+C_2P_2) ], \label{qcasa1} \\
&& \back\back
   {\cal{C}}_{2}^q=\frac {\Ck_{-\m_1\m_3}(2w)}{4w^2}
  \sinh(2wJ)\Sk_{- \m_1\m_3}(2wD)  +
  \frac{\Sk_{-\m_1\m_3}(2w)}{4w}[  P_1C_2-C_1P_2 ]. \label{qcasa2}
\eea
\end{proposition}

We finally remark that the classical $r$--matrix is the same for all 
$U^{(s)}_w g_{(\m_1,+1,\m_3)}$ due to the invariance of the product $wr$
under the transformation (\ref{qgctricks}).

\subsection {The non-standard algebras $U^{(n)}_w
g_{(\m_1,\m_2,+1)}$}  

A similar approach applied to the non-standard $q$--deformation of
$so(2,2)$ leads to:

\begin{proposition}
{\bf (The non-standard quantization of $g_{(\m_1,\m_2,+1)}$)}   If we
keep $\m_3=+1$, then we get a quantum algebra $U^{(n)}_w
g_{(\m_1,\m_2,+1)}$ with $r$--matrix $r= J\wedge P_1  +
D\wedge P_2$ given by 
\bea
&&\back {\Delta P_1 =1 \otimes P_1 + P_1\otimes 1,
   \qquad \Delta P_2 =1 \otimes P_2 + P_2\otimes 1,} \cr
&&\back  {\Delta C_1 = 
   e^{-\tfrac w2 P_2 }\Ck_{-\m_1}(wP_1/2) \otimes C_1  +
   C_1\otimes \Ck_{-\m_1}(wP_1/2) e^{\tfrac w2 P_2 }} \cr 
&& \quad
  +e^{-\tfrac w2 P_2}\Sk_{-\m_1}(wP_1/2) \otimes C_2 
  - C_2 \otimes \Sk_{-\m_1}(wP_1/2) e^{\tfrac w2 P_2 },\cr
&&\back  {\Delta C_2 = 
   e^{-\tfrac w2 P_2 }\Ck_{-\m_1}(wP_1/2) \otimes C_2  +
   C_2\otimes \Ck_{-\m_1}(wP_1/2) e^{\tfrac w2 P_2 }} \cr 
&& \quad
  +e^{-\tfrac w2 P_2}\Sk_{-\m_1}(wP_1/2) \otimes\m_1 C_1 
  - \m_1 C_1 \otimes \Sk_{-\m_1}(wP_1/2) e^{\tfrac w2 P_2 },
\label{coprodgcn}\\
&& \back {\Delta J = 
   e^{-\tfrac w2 P_2 }\Ck_{-\m_1}(wP_1/2) \otimes J  +
   J \otimes \Ck_{-\m_1}(wP_1/2) e^{\tfrac w2 P_2 }} \cr 
&& \quad
  -e^{-\tfrac w2 P_2}\Sk_{-\m_1}(wP_1/2) \otimes\m_1 D
  + \m_1 D \otimes \Sk_{-\m_1}(wP_1/2) e^{\tfrac w2 P_2 },\cr
&&\back  {\Delta D = 
   e^{-\tfrac w2 P_2 }\Ck_{-\m_1}(wP_1/2) \otimes D  +
   D \otimes \Ck_{-\m_1}(wP_1/2) e^{\tfrac w2 P_2 }} \cr 
&& \quad
  -e^{-\tfrac w2 P_2}\Sk_{-\m_1}(wP_1/2) \otimes J
  + J \otimes \Sk_{-\m_1}(wP_1/2) e^{\tfrac w2 P_2 };\nonumber 
\eea
\be
\epsilon(X) =0; \qquad
  \gamma(X) = -e^{{w}  P_2}\ X\ e^{-{w}  P_2}, 
  \qquad X \in \{ J,P_i,C_i,D\} ; 
  \label{antipodgcn}
\ee
\bea
&&[J,P_1]=\frac 2w\sinh(wP_2/2)\Ck_{- \m_1}(wP_1/2),\cr
&&[J,P_2]=\frac 2w\m_1\Sk_{- \m_1}(wP_1/2)\cosh(wP_2/2),\cr
&&[J,C_1]=\frac 12\{C_2,\Ck_{- \m_1}(wP_1/2)\cosh(wP_2/2)\}\cr
&&\qquad\qquad -\frac 12\m_1\{C_1,\Sk_{- \m_1}(wP_1/2)\sinh(wP_2/2)\},\cr
&&[J,C_2]=\frac 12\m_1 \{C_1,\Ck_{- \m_1}(wP_1/2)\cosh(wP_2/2)\}\cr
&&\qquad\qquad -\frac 12\m_1\{C_2,\Sk_{- \m_1}(wP_1/2)\sinh(wP_2/2)\},\cr
&&[P_1,P_2]=[C_1,C_2]=0,\quad [P_1,C_2]=[C_1,P_2]=2\m_2J,
\label{commgcn} \\   
&&[P_1,C_1]=-2\m_2D,\quad [P_2,C_2]=2\m_1\m_2 D,\quad [D,J]=0,\cr
&&[D,P_1]=\frac 2w \Sk_{- \m_1}(wP_1/2)\cosh(wP_2/2),\cr
&&[D,P_2]=\frac 2w\sinh(wP_2/2)\Ck_{- \m_1}(wP_1/2),\cr
&&[D,C_1]=-\frac 12\{C_1,\Ck_{- \m_1}(wP_1/2)\cosh(wP_2/2)\}\cr
&&\qquad\qquad +\frac 12 \{C_2,\Sk_{- \m_1}(wP_1/2)\sinh(wP_2/2)\},\cr
&&[D,C_2]=-\frac 12\{C_2,\Ck_{- \m_1}(wP_1/2)\cosh(wP_2/2)\}\cr
&&\qquad\qquad +\frac 12 \m_1 \{C_1,\Sk_{- \m_1}(wP_1/2)\sinh(wP_2/2)\}.
 \nonumber
\eea
\end{proposition} 

The generators $\langle P_1,P_2,J, D\rangle$ close a Hopf 
subalgebra with deformed
commutation relations. The
invariance of the product
$wr$  under (\ref{qgctrickn}) for $\m_3=+1$ explains again the fact that
the classical $r$--matrix is the same for all $U^{(n)}_w
g_{(\m_1,\m_2,+1)}$.   Central elements  can be
stated as follows:

\begin{proposition}
The center of $U^{(n)}_w g_{(\m_1,\m_2,+1)}$ is generated by
\bea
&& \back \back 
   {\cal{C}}_{1}^q =\m_2 J^2 + \m_1\m_2 D^2 -\frac 1w\m_1\{C_1,
\Sk_{- \m_1}(wP_1/2)\cosh(wP_2/2)\}\cr
&& \back \back 
  \qquad + \frac 1w \{C_2,
\sinh(wP_2/2)\Ck_{- \m_1}(wP_1/2)\}+
\m_1\m_2\Ck_{- \m_1}(wP_1)\cosh(wP_2), \label{qcasb1} 
\eea

\newpage

\bea
&& \back\back 
    {\cal{C}}_{2}^q =  \m_2 J D+
   \frac 1w \Sk_{- \m_1}(wP_1/2)\cosh(wP_2/2)C_2\cr
&& \back\back
   \qquad -\frac 1w C_1 \sinh(wP_2/2)\Ck_{- \m_1}(wP_1/2)
+\frac 12 \m_1\m_2 \Sk_{- \m_1}(wP_1 )\sinh(wP_2 ).
   \label{qcasb2} 
\eea
\end{proposition}

\sect {Non-standard quantum kinematical and conformal algebras}

The non-standard structure $U^{(n)}_w g_{(\m_1,\m_2,+1)}$
 gives rise  to
a new set of physically interesting Hopf algebras.
 Their new properties can be 
highlighted by comparing them,  when possible, to the 
standard ones. This is the case
 for the three upper front edge algebras $g_{(\m_1,+1,+1)}$ 
of Fig.\ 1, with $\m_1=1, 0,
-1$, that are isomorphic to $so(2,2)$, $iso(2,1)$ and $so(3,1)$, 
and   support both
standard and non-standard quantum deformations.

These three algebras have two different realizations at the
classical level. The first one is as isometry algebras of motion groups
of (2+1) Lorentzian spaces with constant curvature (anti-de Sitter,
Minkowski and de Sitter spaces). The second realization 
arises if we consider them as
conformal algebras of flat spaces: the (1+1) Minkowski and Galilean
spaces and the 2d Euclidean plane.

\subsection {A  ``null plane" deformation of (2+1) Poincar\'e and de
Sitter algebras}

The structure of the standard deformation of algebras
$g_{(\mu_1,1,1)}$ is more clearly appreciated in a new  physical 
basis, $\{ H, T_1, T_2, K_1, K_2, L \}$  generating respectively
the time translation, space translations, boosts and space rotation.
The required change of basis from the former one 
$\{P_1, P_2, C_1, C_2, D, J\}$ is
\bea
&& H=\frac{1}{2}(P_2-C_2),\quad 
T_1=\frac{1}{2}(P_2+C_2),\quad T_2=J,\cr
&& K_1=D,\quad   K_2=\frac{1}{2}(C_1-P_1),\quad
 L=\frac{1}{2}(C_1+P_1).\label{5.1}
\eea
We do not give here the standard coproduct nor the deformed
commutators (which can be easily got from (\ref{coprodgcs}--\ref{commgcs}))
 but simply remark
that the standard deformation $U_w^{(s)}g_{(\m_1,1,1)}$ has 
$T_2$ and $K_1$ as primitive generators and was studied in \cite{BHOS3}.

Things are dramatically different for the non-standard
deformation since the primitive generators are $P_1$ and $P_2$,
 therefore, in the basis
(\ref{5.1}) we would find a ``mixture" of primitive 
and non-primitive generators.
Hence,  the most adapted basis to write down the Hopf algebra is 
$\{P_i, C_i, D, J\}$. When these generators are expressed in terms of 
$\{ H, T_i, K_i, L \}$ we get the 
  null plane basis, $\{T_+,T_-, T_2, K_1,  E, F\}$ 
where the new generators $T_+,T_-,E,
F$ are given in terms of the  physical basis  
by the known expressions: 
\be
  T_+=  T_1+H\equiv P_2 ,  \quad T_-=   T_1-H\equiv C_2 , 
\quad E  =  L-K_2\equiv P_1 ,  
\quad F =   L+K_2\equiv C_1 . \label{5.2} 
\ee 
By means of (\ref{5.2})  and after the specialization $\mu_1=0$ and
$\m_2=\mu_3=1$ in formulas (\ref{coprodgcn}--\ref{qcasb2}) we obtain
the coproduct, counit, antipode, deformed Lie brackets and Casimirs
defining the (2+1)   non-standard
$q$--Poincar\'e algebra:
\bea
&&\back\back \Delta  T_+  =1 \otimes  T_+  +  T_+ \otimes 1 ,
   \qquad  \Delta E =1 \otimes E +
E\otimes 1, \cr
&&\back\back  {\Delta  T_-  = 
   e^{-\tfrac w2  T_+  } \otimes  T_-   +
    T_- \otimes  e^{\tfrac w2  T_+  }} ,\quad
  {\Delta  T_2  = 
   e^{-\tfrac w2  T_+  } \otimes  T_2   +
    T_2  \otimes e^{\tfrac w2  T_+  }} ,\label{5.3} \\
&&\back\back {\Delta K_1 = 
   e^{-\tfrac w2  T_+  } \otimes K_1  +
   K_1 \otimes  e^{\tfrac w2  T_+  }} -   \frac w2\left(
   e^{-\tfrac w2  T_+ }E \otimes  T_2 
  +  T_2  \otimes E\, e^{\tfrac w2  T_+  }\right),\cr
&&\back\back  {\Delta F = 
   e^{-\tfrac w2  T_+  }\otimes F  +
   F\otimes  e^{\tfrac w2  T_+  }}  +\frac w2\left(
  e^{-\tfrac w2  T_+ }E \otimes  T_-  
  -  T_-  \otimes E\,e^{\tfrac w2  T_+  }\right);\nonumber
\eea
\be
\epsilon(X) =0; \quad\ 
  \gamma(X) = -e^{{w}   T_+ }\ X\ e^{-{w}   T_+ }, 
  \quad\  X \in \{T_+, T_-,  T_2 , K_1,E,F\} ; 
  \label{5.4}
\ee
\bea
&&[ T_2 ,E]=\frac 2w\sinh(w T_+ /2),\qquad
 [ T_2 ,F]=  T_- \cosh(w T_+ /2) ,\cr
 &&  [E, T_- ]=[F, T_+ ]=2T_2 ,\qquad [E,F]=-2K_1, \cr
 && [K_1 , T_+ ]=\frac 2w\sinh(w T_+ /2),\qquad 
  [K_1 , T_- ]=-  T_- \cosh(w T_+ /2) ,\label{5.5} \\   
&& [K_1 ,E]=E\cosh(w T_+ /2),\cr
&& [K_1 ,F]=-\frac 12\{F,\cosh(w T_+ /2)\} 
+\frac w4 \{ T_- ,E\sinh(w T_+ /2)\},\cr 
&& [ T_2 , T_\pm ]= [ T_+ , T_- ]=[T_2,K_1]=[E, T_+ ]=[F, T_- ]=0; 
\nonumber
\eea
\bea 
 &&  {\cal{C}}_{1}^q = T_2 ^2 + \frac 2w\,  T_- \sinh(w T_+ /2),
\label{5.6} \\
 &&    {\cal{C}}_{2}^q =  T_2 K_1+
    \frac 12 E\, T_-\cosh(w T_+ /2)  -\frac 1w F \sinh(w T_+ /2).
\label{5.7}
\eea

Since the coproduct of $K_1$ involves $E$ and $T_2$, 
the generators $K_1$, $T_+$ and
$T_-$ do not span  a Hopf subalgebra; despite this fact, 
their commutation relations
close a deformed (1+1) Poincar\'e algebra
\cite{Lyak}. On the contrary, the subset 
$\la K_1, T_+ ,   T_2 , E\ra $, whose elements
generate classically the isotopy subalgebra of the null
 plane $x_-=0$ \cite{Lev}, does
span a Hopf subalgebra that can be interpreted as a (1+1)
 $q$--Galilei algebra $\la 
T_+ ,T_2,E \ra $ enlarged with a dilation $\la K_1\ra$.

A similar structure can be readily obtained when 
$\m_1=\pm 1$, leading to the quantum
non-standard de Sitter  algebras. In these cases, 
the curvature of the space-time  
equals to $-\m_1$.

The fact that the kinematical part of the null 
plane description (the isotopy
subalgebra of the null plane) is preserved as a 
Hopf subalgebra under this
non-standard deformation is rather remarkable. 
This fact could be interesting  as a
guide for the physical interpretation of this 
``null plane deformation" of (2+1)
Poincar\'e algebra,  which should be related
to situations where the classical  null plane 
dynamics is relevant (for instance, when
the ultrarrelativistic limit is involved \cite{KogSus}).

\subsection{Conformal algebras in two dimensions}

The Lie algebra of (1+1) affine groups, $\la   J ,  P_1,   P_2  \ra$, with
commutators:
\be
[J , P_1]=  P_2 , \qquad [ J ,  P_2 ]=\m_1  P_1, \qquad [ P_1,  P_2 ]=0, 
\ee
reproduces the Euclidean, Galilei and Poincar\'e algebras for
$\m_1<0, =0, >0$, respectively. All the three algebras can be extended by
adding a dilation generator, $D$, and two special conformal generators,
$C_1,  C_2 $,  which close the algebra $g_{(\m_1,1,1)}$. We therefore get the
second realization above referred to for these three algebras (the
ones allowing both types of deformation) as conformal algebras in two
dimensions.
 
The standard deformation has $J$ and $D$ as primitive 
generators, and
is given by (\ref{coprodgcs}--\ref{commgcs}). We do not
 elaborate upon the comments
made in Sect 4.1.

However, the non-standard deformation is  rather different, 
since now \it
both \rm translations $P_1$ and $P_2$ are \it primitive\rm.
 While in the
``standard" case both pairs $\{P_1,P_2\}$ and $\{ C_1, C_2\}$ 
enter on a completely
symmetrical footing,  this is no longer the case now, because 
 $C_1$ and
$C_2$ are not primitive. The generators $\langle J,P_1,P_2,D\rangle$
span a Hopf subalgebra with
\it deformed \rm commutation rules:
\bea
&&\back {\Delta P_1 =1 \otimes P_1 + P_1\otimes 1,
   \qquad \Delta P_2 =1 \otimes P_2 + P_2\otimes 1,} \cr
&& \back {\Delta J = 
   e^{-\tfrac w2 P_2 }\Ck_{-\m_1}(wP_1/2) \otimes J  +
   J \otimes \Ck_{-\m_1}(wP_1/2) e^{\tfrac w2 P_2 }} \cr 
&& \quad
  -e^{-\tfrac w2 P_2}\Sk_{-\m_1}(wP_1/2) \otimes\m_1 D
  + \m_1 D \otimes \Sk_{-\m_1}(wP_1/2) e^{\tfrac w2 P_2 },\cr
&&\back  {\Delta D = 
   e^{-\tfrac w2 P_2 }\Ck_{-\m_1}(wP_1/2) \otimes D  +
   D \otimes \Ck_{-\m_1}(wP_1/2) e^{\tfrac w2 P_2 }} \cr 
&& \quad
  -e^{-\tfrac w2 P_2}\Sk_{-\m_1}(wP_1/2) \otimes J
  + J \otimes \Sk_{-\m_1}(wP_1/2) e^{\tfrac w2 P_2 };\nonumber 
\eea
\be
\epsilon(X) =0; \qquad
  \gamma(X) = -e^{{w}  P_2}\ X\ e^{-{w}  P_2}, 
  \qquad X \in \{ J,P_i,D\} ; 
\ee
\bea
&&[J,P_1]=\frac 2w\sinh(wP_2/2)\Ck_{- \m_1}(wP_1/2),\cr
&&[J,P_2]=\frac 2w\m_1\Sk_{- \m_1}(wP_1/2)\cosh(wP_2/2),\cr
&&[D,P_1]=\frac 2w \Sk_{- \m_1}(wP_1/2)\cosh(wP_2/2),\cr
&&[D,P_2]=\frac 2w\sinh(wP_2/2)\Ck_{- \m_1}(wP_1/2),\cr
&&[P_1,P_2]=0,\qquad [D,J]=0.
 \nonumber
\eea
 Note that $\langle
J,C_1,C_2,D\rangle$ is not a Hopf subalgebra. 
These properties prompt  to focus
attention to the classical differential realization 
of conformal algebras given by: 
\bea
&&P_1 = \partial_1,  \qquad\qquad    P_2 =\partial_2,\cr
&&  J = -\m_1 x_2\, \partial_1 - x_1 \, \partial_2 + B, \cr
&&D=-  x_1\, \partial_1 - x_2 \, \partial_2 + A + 1,  \\
&&C_1=(x_1^2+\m_1 x_2^2) \, \partial_1 + 2 x_1 x_2 \, \partial_2 -
    2 (A+1) x_1 - 2B x_2, \cr 
&& C_2 =- (x_1^2+\m_1 x_2^2) \,\partial_2 - 2\m_1x_1 x_2 \,\partial_1 
   + 2 \m_1 (A+1) x_2 +  2 B x_1 ,\nonumber
\eea
where $A$ and $B$ are arbitrary real  constants
(for $A=-1$ and $B=0$ this reproduces the fundamental fields for the
local action of the conformal group on the 2d space). The
non-standard $q$--deformed version of this realization gives us:
\bea
&&\!\!\back{P_1=\partial_1,   \qquad\qquad   P_2 =\partial_2,}  \cr
&&\!\!\back{J =-\m_1x_2\frac 2 w \Sk_{-\m_1}(w\partial_1/2)
\cosh(w\partial_2/2) 
    - x_1  \frac 2w \Ck_{-\m_1}(w\partial_1/2)
\sinh(w\partial_2/2)}\cr
&&\quad   + B + \m_1 \Sk_{-\m_1}(w\partial_1/2)
\sinh(w\partial_2/2),\cr
&&\!\!\back{D=-  x_1\frac 2 w \Sk_{-\m_1}(w\partial_1/2)
\cosh(w\partial_2/2)
    - x_2 \frac 2 w \Ck_{-\m_1}(w\partial_1/2)
\sinh(w\partial_2/2)}\cr
&&\quad + A +\Ck_{-\m_1}(w\partial_1/2)\cosh(w\partial_2/2),
 \label{5.10}   \\
&&\!\!\back{C_1=(x_1^2+\m_1 x_2^2)\frac 2 w
\Sk_{-\m_1}(w\partial_1/2)\cosh(w\partial_2/2) +  x_1 x_2\frac 4
w\Ck_{-\m_1}(w\partial_1/2)\sinh(w\partial_2/2)}\cr 
&&\quad  -2 x_1  [A+
\Ck_{-\m_1}(w\partial_1/2)\cosh(w\partial_2/2)]  - 2 x_2 [B + \m_1
\Sk_{-\m_1}(w\partial_1/2)\sinh(w\partial_2/2)] ,\cr 
&&\!\!\back{C_2 =- (x_1^2+\m_1 x_2^2)
\frac 2 w \Ck_{-\m_1}(w\partial_1/2) \sinh(w\partial_2/2)
 -   \m_1 x_1 x_2 \frac 4 w \Sk_{-\m_1}(w\partial_1/2)
\cosh(w\partial_2/2)}\cr
&&\quad+ 2\m_1  x_2
  [A+\Ck_{-\m_1}(w\partial_1/2)\cosh(w\partial_2/2)] + 
 2   x_1   [B +
\m_1\Sk_{-\m_1}(w\partial_1/2)\sinh(w\partial_2/2)].
\nonumber 
\eea
 Note also that the
symmetric $q$--derivative
\bea
D_qf(x):= \frac{\sinh (w\partial_x/2)}{w/2} f(x) &=&
   \frac{\exp{(w\partial_x/2)}-\exp{(-w\partial_x/2)}}{ w}  f(x)\cr
 &=&
   \frac{f(x+w/2)-f(x-w/2)}{ w}   
\eea
is naturally contained in the realization (\ref{5.10}) for both
$P_1$ and $  P_2 $ generators.

 We recall that within the known
standard deformations of (1+1) and (2+1) algebras, 
such a discretization
appears only in {\it one} spatial direction 
\cite{CGST1,CGST2}. Therefore a ``conformal"
approach to this problem seems to be promising 
as it would allow  a kind of  complete
discretization of the  space-time in (1+1) dimensions.

\sect {Concluding remarks}

As a general result, we emphasize that a systematic
 use of the theory of
graded contractions provides a well defined and
 encompassing framework to
study $q$--deforma\-tions of some real non-semisimple 
algebras in a
straightforward way. Both standard and non-standard 
deformations of $so(2,2)$ that we
have introduced in this paper generate by contraction  
 quantum deformations
of some relevant kinematical and conformal groups. Some 
of them are new and others
coincide with   already known quantum algebras as 
the $q$--Poincar\'e and $q$--de Sitter
algebras obtained in ref. \cite{BHOS3}.

A  point worth stressing is that, in general, the contraction 
process needs a
careful examination; for each $so(2,2)$ deformation not all possible
classical graded contractions induce quantum deformations for the
contracted algebras. The same analysis can be performed for the quantum
$R$--matrices. This fact is related to some problems which arise 
 when contractions of
quantum groups are made in a  naive way.

By using consistently only real forms some
contraction processes  found in the literature can be also
 clarified. Of course, if
complex coefficients are allowed in the changes of basis, 
such as it has been done in 
\cite{CGST1,CGST2}, then further possibilities are opened. 
In this way and starting from
any standard quantum $iso(2,1)$ algebra we could get the 
(2+1) $\k$--Poincar\'e
\cite{Gill,Luk}, which does not appear as such in our scheme. 
Another example along this line is provided by the
contraction sequence $SO(4)_q\to E(3)_q\to G(2)_q$ studied in \cite{CGST1,CGST2}, that  
corresponds, in our context of real forms,  to the standard quantum algebras
$so(2,2)_q\to iso(2,1)_q\to i'iso(1,1)_q$, this is,   $U_w^{(s)}g(1,1,1)\to
U_w^{(s)}g(0,1,1)\to  U_w^{(s)}g(0,1,0)$. 
In this sense, it is important to recall that as a real form, the  algebra
$i'iso(1,1)_q$ is not isomorphic to the (2+1) Galilean algebra: the latter {does
not} include a central generator while the algebra  $i'iso(1,1)$ does.

A way opened by this paper would consist in the construction  
 of the quantum
groups corresponding to the non-standard family of algebras. For all of 
them, the
existence of a star product that quantizes their classical Poisson--Lie
 structures is
guaranteed (their $r$--matrices exist and are degenerate). Moreover a
solution of the quantum YBE linked to each non-standard quantum algebra 
can be obtained
as a biproduct of the bidifferential operator that defines the star 
product on the
 group  \cite{Tak}. Also, 
the natural link of the non-standard
deformation of the Poincar\'e algebra to the null plane basis and 
its possible
physical interpretation should require further study. Should this 
deformation survive for
the (3+1) case, we would get  a new quantum (3+1) Poincar\'e 
algebra whose
features would   be certainly different from the known 
$\k$--Poincar\'e algebra
\cite{Gill,Luk}.  Work on these lines is currently in progress.

\bigskip\medskip

\noindent
{\large{{\bf Acknowledgements}}}

\bigskip
This work has been partially supported by DGICYT de Espa\~na (Project 
PB91--0196).

\end{document}